\documentclass[lettersize,journal]{IEEEtran}
\usepackage{amsmath,amsfonts}
\usepackage{algorithmic}
\usepackage{algorithm}
\usepackage{array}
\usepackage{textcomp}
\usepackage{stfloats}
\usepackage{url}
\usepackage{verbatim}
\usepackage{graphicx}
\usepackage{cite}

\usepackage[utf8]{inputenc} 
\usepackage[T1]{fontenc}    
\usepackage{hyperref}       
\usepackage{url}            
\usepackage{booktabs}       
\usepackage{amsfonts}       
\usepackage{nicefrac}       
\usepackage{float}

\usepackage{microtype}      
\usepackage{xcolor}         
\usepackage{multirow}
\usepackage{graphicx}
\usepackage{subcaption}
\usepackage{pifont}    
\usepackage{tcolorbox}

\usepackage{musicography}
\usepackage{tikz}
\usepackage{multicol}
\usepackage{enumitem}

\definecolor{MyLightBlue}{HTML}{0072BD}
\definecolor{MyLightPurple}{HTML}{7E2F8E}
\definecolor{MyLightRed}{HTML}{D95319}
\definecolor{MyLightGreen}{HTML}{77AC30}

\definecolor{SystemBackBase}{HTML}{F5F5F5}
\definecolor{SystemFrame}{HTML}{666666}
\definecolor{UserBackBase}{HTML}{DAE8FC}
\definecolor{UserFrame}{HTML}{6C8EBF}
\definecolor{LLMBackBase}{HTML}{E1D5E7}
\definecolor{LLMFrame}{HTML}{9673A6}

\colorlet{SystemBack}{SystemBackBase!10}
\colorlet{UserBack}{UserBackBase!10}
\colorlet{LLMBack}{LLMBackBase!10}

\begin{document}




\title{CapSpeech: Enabling Downstream Applications in \\Style-Captioned Text-to-Speech}

\author{Helin Wang$^{\dagger}$, Jiarui Hai$^{\dagger}$, Dading Chong, Karan Thakkar, Tiantian Feng, Dongchao Yang, Junhyeok Lee,\\Thomas Thebaud, Laureano Moro-Velázquez, Jesús Villalba, Zengyi Qin, Shrikanth Narayanan,~\IEEEmembership{Fellow,~IEEE}, \\Mounya Elhiali,~\IEEEmembership{Senior~Member,~IEEE} and Najim Dehak,~\IEEEmembership{Senior~Member,~IEEE}
\thanks{Helin Wang and Jiarui Hai are co-first authors.}
\thanks{
Helin Wang, Jiarui Hai, Karan Thakkar, Junhyeok Lee, Thomas Thebaud, Laureano Moro-Velázquez, Jesús Villalba, Mounya Elhiali and Najim Dehak are with the Johns
Hopkins University (email: $\{$hwang258, jhai2$\}$@jhu.edu).}
\thanks{
Dading Chong is with the Peking University.
Dongchao Yang is with the Chinese University of Hong Kong. Tiantian Feng and Shrikanth Narayanan are with the University of Southern California. Zengyi Qin is with the Massachusetts Institute of Technology.
}}

\maketitle

\begin{abstract}
Recent advancements in generative artificial intelligence have significantly transformed the field of style-captioned text-to-speech synthesis (CapTTS). However, adapting CapTTS to real-world applications remains challenging due to the lack of standardized, comprehensive datasets and limited research on downstream tasks built upon CapTTS.
To address these gaps, we introduce \textbf{CapSpeech}, a new benchmark designed for a series of CapTTS-related tasks, including style-captioned text-to-speech synthesis with sound events (\textbf{CapTTS-SE}), accent-captioned TTS (\textbf{AccCapTTS}), emotion-captioned TTS (\textbf{EmoCapTTS}), and text-to-speech synthesis for chat agent (\textbf{AgentTTS}).
CapSpeech comprises over 10 million machine-annotated audio-caption pairs and nearly 0.36 million human-annotated audio-caption pairs.
In addition, we introduce \textbf{two new datasets} collected and recorded by a professional voice actor and experienced audio engineers, specifically for the AgentTTS and CapTTS-SE tasks.
Alongside the datasets, we conduct comprehensive experiments using both autoregressive and non-autoregressive models on CapSpeech.
Our results demonstrate high-fidelity and highly intelligible speech synthesis across a diverse range of speaking styles.
To the best of our knowledge, CapSpeech is the largest available dataset offering comprehensive annotations for CapTTS-related tasks. The experiments and findings further provide valuable insights into the challenges of developing CapTTS systems.
\end{abstract}


\section{Introduction}
In recent years, large-scale text-to-speech (TTS) synthesis has witnessed remarkable progress, exemplified by models such as VALL-E 2 \cite{DBLP:journals/corr/abs-2406-05370}, NaturalSpeech 3 \cite{DBLP:conf/icml/JuWS0XYLLST000024}, LLasa \cite{DBLP:journals/corr/abs-2502-04128}, SimpleSpeech 2 \cite{DBLP:journals/corr/abs-2408-13893}, CosyVoice 2 \cite{DBLP:journals/corr/abs-2412-10117}, and MaskGCT \cite{DBLP:journals/corr/abs-2409-00750}. Most of these efforts have centered on modeling audio-based speaker characteristics, such as zero-shot TTS with audio prompts \cite{DBLP:journals/corr/abs-2406-02430,DBLP:journals/corr/abs-2410-06885,DBLP:conf/slt/EskimezWTLTXYZTTLZK24}, category-based traits \cite{DBLP:conf/slt/WuWETTTLXZLK24,DBLP:conf/icassp/GuoDCY23,DBLP:journals/taslp/LeiYWX22,DBLP:journals/taslp/ZhouZZWL24}, or embedding-based speaker representations \cite{DBLP:journals/corr/abs-2406-07803,DBLP:conf/icassp/Tang000W24}. However, the detailed understanding of speaking style from audio, particularly its subtle nuances, has received limited attention.

The style of speech encompasses both intrinsic traits tied to a speaker’s identity (e.g., age, gender, timbre) and expressive style traits specific to individual utterances (e.g., emotion, speaking rate). Recent studies have introduced the use of natural language captions to describe these stylistic elements—a paradigm referred to as prompt TTS, expressive TTS, or style-captioned TTS \cite{DBLP:conf/icassp/GuoLWZT23,DBLP:conf/mm/Jin0W0ZZQ024,DBLP:conf/iclr/LengGSJ0LLYZS0024,DBLP:conf/icassp/ShimizuYKSDKT24,DBLP:conf/interspeech/NguyenHDSGFRCSH23,DBLP:journals/taslp/YangLHWM24,DBLP:conf/mm/ZhuTWHX00Z024}. In this work, we adopt the term style-captioned TTS and abbreviate it as CapTTS.

Developing a CapTTS system necessitates a large corpus of audio-caption pairs accompanied by transcriptions, the annotation of which is labor-intensive and costly. Parler-TTS \cite{DBLP:journals/corr/abs-2402-01912,lacombe-etal-2024-parler-tts} addressed this challenge by automatically annotating basic speech style attributes such as pitch and speed using signal processing tools. ParaSpeechCaps \cite{DBLP:journals/corr/abs-2503-04713} collected speaker-level intrinsic tags and utterance-level situational tags, LibriTTS-P \cite{DBLP:journals/corr/abs-2406-07969} offered speaker identity-based captions, and EmoVoice-DB \cite{yang2025emovoice} provided emotion-based captions.
However, these existing datasets lack a unified and comprehensive framework for style captioning, making cross-domain comparisons difficult. Moreover, there has been limited exploration of downstream applications, such as transferring models to new caption styles or incorporating sound events into the synthesized speech.

To address the challenges outlined above, we present CapSpeech, a novel benchmark featuring standardized and comprehensive datasets for CapTTS and its related downstream tasks. 
CapSpeech comprises a pretraining stage with large-scale captioned speech,
as well as five downstream tasks: CapTTS, CapTTS-SE, accent-captioned TTS (AccCapTTS), emotion-captioned TTS (EmoCapTTS), and TTS for chat agent (AgentTTS).
The pretraining datasets consist of over 10 million machine-annotated audio-caption pairs, while the downstream datasets contain 358,783 human-annotated audio-caption pairs. These datasets encompass a broad range of intrinsic speaker traits and expressive style traits, curated from a wide array of audio sources, including Emilia \cite{DBLP:conf/slt/HeSWLGHLYLSWCZW24}, GigaSpeech \cite{DBLP:conf/interspeech/ChenCWDZWSPTZJK21}, CommonVoice \cite{DBLP:conf/lrec/ArdilaBDKMHMSTW20}, MLS \cite{DBLP:conf/interspeech/PratapXSSC20}, LibriTTS-R \cite{DBLP:conf/interspeech/KoizumiZKDYMB0H23}, VoxCeleb \cite{DBLP:conf/interspeech/NagraniCZ17}, VoxCeleb2 \cite{DBLP:conf/interspeech/ChungNZ18}, EARS \cite{DBLP:journals/corr/abs-2406-06185}, Expresso \cite{DBLP:conf/interspeech/NguyenHDSGFRCSH23}, VCTK \cite{Yamagishi2019VCTK}, VGGSound \cite{DBLP:conf/icassp/ChenXVZ20}, FSDKaggle2018 \cite{DBLP:conf/dcase/FonsecaPFEFPS18}, and ESC-50 \cite{piczak2015dataset}.
In addition, we introduce two new datasets: one for AgentTTS, built using professionally recorded voice actor speech, and another for CapTTS-SE, processed by five experienced audio engineers.

In addition, we develop two style-captioned TTS models based on state-of-the-art generative TTS backbones: one autoregressive (AR) and one non-autoregressive (NAR). We evaluate these models on the CapSpeech benchmark across 5 downstream tasks. For a comprehensive evaluation, both objective and subjective metrics are employed to assess speech style consistency, audio quality, and intelligibility.

In summary, our contributions are as follows: 

\begin{itemize}
    \item We introduce CapSpeech, a \textbf{large-scale} style-captioned benchmark for CapTTS that encompasses diverse speech styles. 
    \item We propose \textbf{5 CapTTS-related downstream tasks, 4 of which are new}, aimed at practical CapTTS applications, 
introduce \textbf{9,353,823 new captions} for existing speech corpora, and curate \textbf{3 new speech datasets} specifically designed for the CapTTS-SE and AgentTTS tasks, thereby enhancing the benchmark’s coverage of real-world scenarios.
\item We train and evaluate two CapTTS models, \textbf{CapSpeech-AR and CapSpeech-NAR}, which address the lack of benchmarks for CapTTS downstream tasks, demonstrate the effectiveness of the proposed dataset, and highlight key challenges and future research directions.
\item We \textbf{publicly release our datasets, listening samples, source code, pretrained checkpoints, and evaluation toolkit} to support future research\footnote{\href{https://github.com/WangHelin1997/CapSpeech}{https://github.com/WangHelin1997/CapSpeech}}. All resources are released under the CC BY-NC 4.0 license (Creative Commons Attribution-NonCommercial), which permits use for non-commercial research purposes with appropriate attribution.
\end{itemize}

\begin{table*}[t]
  \caption{A comparison of English speech style-captioned datasets. I1–I5 denote intrinsic speaker traits: age (I1), gender (I2), timbre (I3), mean pitch (I4), and accent (I5). E1–E4 represent expressive style traits: speaking rate (E1), emotion (E2), expressiveness of tone (E3), and volume (E4).}
  \label{tab:data1}
  \vskip 0.05in
  \scriptsize
  \setlength{\tabcolsep}{6.5pt}  
  \centering
  \begin{tabular}{l|ccccc|cccc|ccc}
    \toprule
    \multirow{2}{*}{\textbf{Dataset}} & \multicolumn{5}{c|}{\textbf{Intrinsic Traits}}& \multicolumn{4}{c|}{\textbf{Expressive Traits}} & \multicolumn{3}{c}{\textbf{Coverage \& Access}}  \\
        \cmidrule(lr){2-7} \cmidrule(lr){7-11} \cmidrule(lr){11-13} 
    & I1 & I2 & I3 & I4 & I5 & E1 & E2 & E3 & E4 & Duration (h) & Sound Event & Open Source \\
        \midrule
    PromptSpeech \cite{DBLP:conf/icassp/GuoLWZT23} &\textcolor{MyLightRed}{\ding{55}} & \textcolor{MyLightGreen}{\ding{51}} &\textcolor{MyLightRed}{\ding{55}} & \textcolor{MyLightGreen}{\ding{51}} &\textcolor{MyLightRed}{\ding{55}}& \textcolor{MyLightGreen}{\ding{51}}& \textcolor{MyLightGreen}{\ding{51}}&\textcolor{MyLightRed}{\ding{55}}& \textcolor{MyLightGreen}{\ding{51}} & 0.3k &\textcolor{MyLightRed}{\ding{55}}& \textcolor{MyLightGreen}{\ding{51}}\\
    Expresso \cite{DBLP:conf/interspeech/NguyenHDSGFRCSH23} &\textcolor{MyLightRed}{\ding{55}} &\textcolor{MyLightRed}{\ding{55}} &\textcolor{MyLightRed}{\ding{55}} &\textcolor{MyLightRed}{\ding{55}} &\textcolor{MyLightRed}{\ding{55}}&\textcolor{MyLightRed}{\ding{55}}&\textcolor{MyLightGreen}{\ding{51}} & \textcolor{MyLightGreen}{\ding{51}} &\textcolor{MyLightRed}{\ding{55}} & 47 &\textcolor{MyLightRed}{\ding{55}}& \textcolor{MyLightGreen}{\ding{51}}\\
    EARS \cite{DBLP:journals/corr/abs-2406-06185} & \textcolor{MyLightGreen}{\ding{51}}& \textcolor{MyLightGreen}{\ding{51}}&\textcolor{MyLightRed}{\ding{55}}& \textcolor{MyLightGreen}{\ding{51}}& \textcolor{MyLightGreen}{\ding{51}}& \textcolor{MyLightGreen}{\ding{51}}& \textcolor{MyLightGreen}{\ding{51}}& \textcolor{MyLightGreen}{\ding{51}} & \textcolor{MyLightGreen}{\ding{51}} & 60 &\textcolor{MyLightRed}{\ding{55}}& \textcolor{MyLightGreen}{\ding{51}}\\
    TextrolSpeech \cite{DBLP:conf/icassp/JiZ00CDHZ24} &\textcolor{MyLightRed}{\ding{55}} & \textcolor{MyLightGreen}{\ding{51}}&\textcolor{MyLightRed}{\ding{55}} & \textcolor{MyLightGreen}{\ding{51}}&\textcolor{MyLightRed}{\ding{55}}& \textcolor{MyLightGreen}{\ding{51}}  & \textcolor{MyLightGreen}{\ding{51}}&\textcolor{MyLightRed}{\ding{55}}& \textcolor{MyLightGreen}{\ding{51}}& 0.3k &\textcolor{MyLightRed}{\ding{55}}& \textcolor{MyLightGreen}{\ding{51}}\\
    SpeechCraft \cite{DBLP:conf/mm/Jin0W0ZZQ024} & \textcolor{MyLightGreen}{\ding{51}}& \textcolor{MyLightGreen}{\ding{51}}&\textcolor{MyLightRed}{\ding{55}} & \textcolor{MyLightGreen}{\ding{51}} &\textcolor{MyLightRed}{\ding{55}}& \textcolor{MyLightGreen}{\ding{51}} & \textcolor{MyLightGreen}{\ding{51}}&\textcolor{MyLightRed}{\ding{55}}& \textcolor{MyLightGreen}{\ding{51}} & 2.4k &\textcolor{MyLightRed}{\ding{55}}& \textcolor{MyLightGreen}{\ding{51}}\\
    VccmDataset \cite{DBLP:journals/corr/abs-2406-01205}&\textcolor{MyLightRed}{\ding{55}} & \textcolor{MyLightGreen}{\ding{51}}&\textcolor{MyLightRed}{\ding{55}} & \textcolor{MyLightGreen}{\ding{51}}&\textcolor{MyLightRed}{\ding{55}}& \textcolor{MyLightGreen}{\ding{51}}  & \textcolor{MyLightGreen}{\ding{51}}&\textcolor{MyLightRed}{\ding{55}}& \textcolor{MyLightGreen}{\ding{51}}& 0.3k &\textcolor{MyLightRed}{\ding{55}}& \textcolor{MyLightGreen}{\ding{51}}\\
    LibriTTS-P \cite{DBLP:journals/corr/abs-2406-07969} &\textcolor{MyLightGreen}{\ding{51}} &\textcolor{MyLightGreen}{\ding{51}}&\textcolor{MyLightGreen}{\ding{51}}&\textcolor{MyLightGreen}{\ding{51}}&\textcolor{MyLightRed}{\ding{55}}&\textcolor{MyLightGreen}{\ding{51}}&\textcolor{MyLightRed}{\ding{55}} & \textcolor{MyLightGreen}{\ding{51}}& \textcolor{MyLightGreen}{\ding{51}}& 0.6k &\textcolor{MyLightRed}{\ding{55}} & \textcolor{MyLightGreen}{\ding{51}} \\
    DreamVoiceDB \cite{hai2024dreamvoice} &\textcolor{MyLightGreen}{\ding{51}} &\textcolor{MyLightGreen}{\ding{51}} &\textcolor{MyLightGreen}{\ding{51}} &\textcolor{MyLightRed}{\ding{55}} &\textcolor{MyLightRed}{\ding{55}} &\textcolor{MyLightRed}{\ding{55}} &\textcolor{MyLightRed}{\ding{55}} & \textcolor{MyLightGreen}{\ding{51}} &\textcolor{MyLightRed}{\ding{55}} & 0.3k &\textcolor{MyLightRed}{\ding{55}} & \textcolor{MyLightGreen}{\ding{51}} \\
    ParlerTTS \cite{DBLP:journals/corr/abs-2402-01912} & \textcolor{MyLightRed}{\ding{55}}& \textcolor{MyLightGreen}{\ding{51}} & \textcolor{MyLightRed}{\ding{55}} & \textcolor{MyLightGreen}{\ding{51}} & \textcolor{MyLightGreen}{\ding{51}}& \textcolor{MyLightGreen}{\ding{51}}  &\textcolor{MyLightRed}{\ding{55}} & \textcolor{MyLightGreen}{\ding{51}} &\textcolor{MyLightRed}{\ding{55}} & 45k &\textcolor{MyLightRed}{\ding{55}} & \textcolor{MyLightGreen}{\ding{51}}\\
    ParaSpeechCaps \cite{DBLP:journals/corr/abs-2503-04713} &\textcolor{MyLightRed}{\ding{55}} & \textcolor{MyLightGreen}{\ding{51}} & \textcolor{MyLightGreen}{\ding{51}} & \textcolor{MyLightGreen}{\ding{51}}& \textcolor{MyLightGreen}{\ding{51}}& \textcolor{MyLightGreen}{\ding{51}}& \textcolor{MyLightGreen}{\ding{51}}& \textcolor{MyLightGreen}{\ding{51}}& \textcolor{MyLightGreen}{\ding{51}} &  2.9k &\textcolor{MyLightRed}{\ding{55}} & \textcolor{MyLightGreen}{\ding{51}}\\
    NonVerbalSpeech-38K \cite{ye2025scalable}&\textcolor{MyLightRed}{\ding{55}}&\textcolor{MyLightRed}{\ding{55}}&\textcolor{MyLightRed}{\ding{55}}&\textcolor{MyLightRed}{\ding{55}}&\textcolor{MyLightRed}{\ding{55}}&\textcolor{MyLightRed}{\ding{55}}&\textcolor{MyLightRed}{\ding{55}}&\textcolor{MyLightRed}{\ding{55}}&\textcolor{MyLightRed}{\ding{55}}&0.1k&\textcolor{MyLightGreen}{\ding{51}}&\textcolor{MyLightGreen}{\ding{51}}\\
    
    PromptTTS2 \cite{DBLP:conf/iclr/LengGSJ0LLYZS0024} &\textcolor{MyLightRed}{\ding{55}}& \textcolor{MyLightGreen}{\ding{51}} & \textcolor{MyLightRed}{\ding{55}} & \textcolor{MyLightGreen}{\ding{51}} &\textcolor{MyLightRed}{\ding{55}}& \textcolor{MyLightGreen}{\ding{51}} &\textcolor{MyLightRed}{\ding{55}}&\textcolor{MyLightRed}{\ding{55}}& \textcolor{MyLightGreen}{\ding{51}} & 44k&\textcolor{MyLightRed}{\ding{55}}&\textcolor{MyLightRed}{\ding{55}}\\
    VoxInstruct \cite{DBLP:conf/mm/0002QJZLZ0024}& \textcolor{MyLightGreen}{\ding{51}}& \textcolor{MyLightGreen}{\ding{51}}&\textcolor{MyLightRed}{\ding{55}} & \textcolor{MyLightGreen}{\ding{51}} &\textcolor{MyLightRed}{\ding{55}}& \textcolor{MyLightGreen}{\ding{51}} & \textcolor{MyLightGreen}{\ding{51}}&\textcolor{MyLightRed}{\ding{55}}& \textcolor{MyLightGreen}{\ding{51}} & 1.5k &\textcolor{MyLightRed}{\ding{55}}&\textcolor{MyLightRed}{\ding{55}}\\
    FleSpeech \cite{DBLP:journals/corr/abs-2501-04644} & \textcolor{MyLightRed}{\ding{55}}& \textcolor{MyLightGreen}{\ding{51}} & \textcolor{MyLightRed}{\ding{55}} & \textcolor{MyLightGreen}{\ding{51}} &\textcolor{MyLightRed}{\ding{55}}& \textcolor{MyLightGreen}{\ding{51}}  & \textcolor{MyLightGreen}{\ding{51}}  & \textcolor{MyLightGreen}{\ding{51}} & \textcolor{MyLightGreen}{\ding{51}} & 0.6k &\textcolor{MyLightRed}{\ding{55}} &\textcolor{MyLightRed}{\ding{55}}\\
    
    Audiobox \cite{DBLP:journals/corr/abs-2312-15821} & \textcolor{MyLightGreen}{\ding{51}} & \textcolor{MyLightGreen}{\ding{51}}&\textcolor{MyLightRed}{\ding{55}} & \textcolor{MyLightGreen}{\ding{51}} & \textcolor{MyLightGreen}{\ding{51}}& \textcolor{MyLightGreen}{\ding{51}} & \textcolor{MyLightGreen}{\ding{51}}&\textcolor{MyLightRed}{\ding{55}}&\textcolor{MyLightRed}{\ding{55}} & >0.5k & \textcolor{MyLightGreen}{\ding{51}} &\textcolor{MyLightRed}{\ding{55}}\\
    \midrule
    \textbf{CapSpeech} &\textcolor{MyLightGreen}{\ding{51}}&\textcolor{MyLightGreen}{\ding{51}}&\textcolor{MyLightGreen}{\ding{51}}&\textcolor{MyLightGreen}{\ding{51}}&\textcolor{MyLightGreen}{\ding{51}}&\textcolor{MyLightGreen}{\ding{51}} & \textcolor{MyLightGreen}{\ding{51}} & \textcolor{MyLightGreen}{\ding{51}} &\textcolor{MyLightGreen}{\ding{51}} & 33.6k &\textcolor{MyLightGreen}{\ding{51}}&\textcolor{MyLightGreen}{\ding{51}} \\
    \bottomrule
  \end{tabular}
\end{table*}

\section{Related Works}
\subsection{Speech Style-Captioned Datasets}
Speech datasets with natural language style captions offer greater flexibility and controllability compared to those that rely on audio prompts.
Early efforts such as NLSpeech \cite{DBLP:journals/taslp/YangLHWM24} recruited human annotators to describe speech emotions, laying the groundwork for later datasets like PromptStyle \cite{DBLP:conf/interspeech/LiuZLCW0L23} and MM-TTS \cite{DBLP:conf/aaai/GuanLLHWLHLH24}.
TextrolSpeech \cite{DBLP:conf/icassp/JiZ00CDHZ24} aggregates several existing emotion datasets, primarily focusing on emotional states and a few basic style tags.
In contrast, Expresso \cite{DBLP:conf/interspeech/NguyenHDSGFRCSH23} and EARS \cite{DBLP:journals/corr/abs-2406-06185} broaden the scope by covering a wider range of spontaneous expressive styles.
Meanwhile, datasets such as LibriTTS-P \cite{DBLP:journals/corr/abs-2406-07969} and DreamVoiceDB \cite{hai2024dreamvoice} focus on intrinsic speaker traits, providing captioned annotations for LibriTTS-R \cite{DBLP:conf/interspeech/KoizumiZKDYMB0H23},
while Coco-Nut \cite{DBLP:conf/asru/WatanabeTSNXS23,DBLP:journals/corr/abs-2403-13353} captures subjective descriptions of voice characteristics beyond traditional acoustic parameters.
More recently, ParaSpeechCaps \cite{DBLP:journals/corr/abs-2503-04713} introduces a human-annotated dataset that covers a richer and more diverse set of speaking style attributes.
Given the high cost and limited scalability of human annotation, several works have explored automatic or semi-automatic approaches to expand style-captioned speech datasets.
PromptSpeech \cite{DBLP:conf/icassp/GuoLWZT23} and PromptTTS2 \cite{DBLP:conf/iclr/LengGSJ0LLYZS0024} synthesize speech with diverse speaker identities and emotions using commercial TTS systems.
Parler-TTS \cite{DBLP:journals/corr/abs-2402-01912,lacombe-etal-2024-parler-tts} proposes large-scale annotation of basic style tags (e.g., pitch, speed) by leveraging signal processing techniques and rule-based binning.
Other efforts, such as FleSpeech \cite{DBLP:journals/corr/abs-2501-04644} and VccmDataset \cite{DBLP:journals/corr/abs-2406-01205}, re-caption existing speech datasets like TextrolSpeech using large language models to improve label quality and expressiveness.
VoxInstruct \cite{DBLP:conf/mm/0002QJZLZ0024} and SpeechCraft \cite{DBLP:conf/mm/Jin0W0ZZQ024} further extend this direction by introducing more fine-grained emotion tags.
In addition, AudioBox \cite{DBLP:journals/corr/abs-2312-15821} integrates multiple strategies, combining large-scale automatically generated basic tags with high-quality human-annotated rich style labels to build a scalable and diverse dataset. It also supports sound events such as laughter and clapping, which are important for expressive speech generation.

Table~\ref{tab:data1} summarizes existing English style-captioned datasets. 
To the best of our knowledge, the proposed CapSpeech is the \textbf{first large-scale open-source dataset} that covers a wide range of both intrinsic speaker traits and expressive speaking styles. Similar to AudioBox \cite{DBLP:journals/corr/abs-2312-15821} and NonVerbalSpeech-38K \cite{ye2025scalable}, CapSpeech also supports speech generation with sound events.

\subsection{Speech Style-captioned TTS Models}
Most existing methods for speech style-captioned TTS build upon prior audio-prompted TTS frameworks. For example, PromptTTS \cite{DBLP:conf/icassp/GuoLWZT23} is developed on top of FastSpeech 2 \cite{DBLP:conf/iclr/0006H0QZZL21}, while PromptTTS2 \cite{DBLP:conf/iclr/LengGSJ0LLYZS0024} is based on NaturalSpeech 2 \cite{DBLP:conf/iclr/ShenJ0LL00Z024}. Similarly, Salle \cite{DBLP:conf/icassp/JiZ00CDHZ24} builds on VALL-E \cite{DBLP:journals/corr/abs-2301-02111}, and Parler-TTS \cite{DBLP:journals/corr/abs-2402-01912} further introduces cross-attention mechanisms to better integrate caption features.
In parallel, diffusion-based approaches such as InstructTTS \cite{DBLP:journals/taslp/YangLHWM24} and AudioBox \cite{DBLP:journals/corr/abs-2312-15821} have also emerged, offering alternative generative paradigms for style-controllable speech synthesis.
In this work, we adopt both autoregressive and non-autoregressive models to evaluate performance on our proposed benchmark.

\section{CapSpeech Benchmark}
CapSpeech comprises a pretraining (PT) stage that leverages large-scale machine-annotated speech–caption pairs, followed by a Supervised Fine-Tuning (SFT) stage using human-annotated captioned speech. In this section, we introduce the tasks and datasets included in the CapSpeech.

\subsection{Tasks and Applications}
As shown in Fig.~\ref{fig:tasks}, the CapSpeech benchmark includes the following five tasks:

\begin{figure*}
  \centering
  \includegraphics[width=0.8\linewidth]{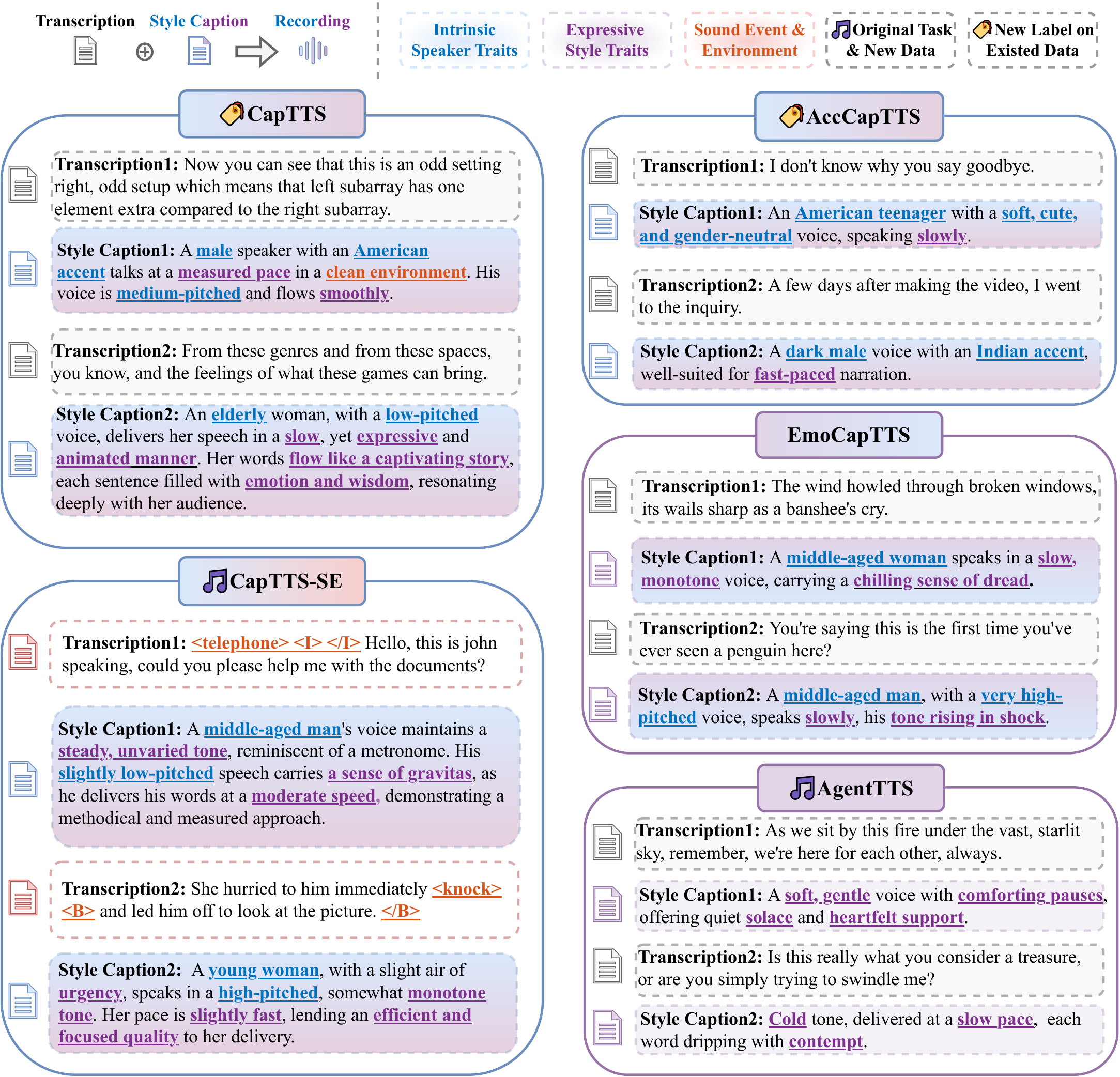}
\caption{Overview of the proposed CapSpeech benchmark and its applications.
    Here, \textcolor{MyLightBlue}{\textbf{\underline{blue}}} words denote intrinsic speaker traits, \textcolor{MyLightPurple}{\textbf{\underline{purple}}} words indicate expressive style traits, and \textcolor{MyLightRed}{\textbf{\underline{red}}} words represent environment description and special sound event tokens used in CapTTS-SE transcriptions.}
  \label{fig:tasks}
\vspace{-1em}
\end{figure*}

\textbf{CapTTS} synthesizes speech from text input conditioned on a natural language style caption that describes desired attributes of the output—such as speaker traits (e.g., age, gender, accent), expressive styles (e.g., emotion, speaking rate), or situational context (e.g., conversational tone, whispering).
This subtask reflects general use cases of caption-based TTS. While it is less focused on specific applications, it provides a versatile benchmark for generating speech under diverse, caption-guided conditions.

\textbf{CapTTS-SE} extends CapTTS by allowing the synthesis of speech that includes non-verbal sound events. The speaking style is specified through a natural language caption, while sound events (e.g., door knocking, applause, dog barking) are explicitly indicated in the transcription. Sound events can either play in the background of the speech or be inserted at specific points within the utterance. This subtask supports applications like audiobooks and livestreaming—any scenario where sound events enhance the experience beyond vanilla speech-only TTS.

\textbf{AccCapTTS} focuses specifically on accent control, in contrast to the broader functionality addressed by CapTTS. Unlike traditional accent-conditioned TTS systems that rely on predefined categories, AccCapTTS offers more user-friendly and flexible control through free-form natural language prompts. This subtask enables applications such as cross-cultural voice design, personalized speech synthesis, and localized content creation—scenarios where nuanced accent control greatly enhances realism and user engagement.

\textbf{EmoCapTTS} generates speech from text input conditioned on a natural language caption that describes both the speaker's emotional state and identity. Unlike traditional emotion TTS systems that rely on discrete categories (e.g., happy, sad, angry)~\cite{DBLP:conf/icassp/Tang000W24,DBLP:conf/interspeech/Tang0W0023}, EmoCapTTS enables more flexible and expressive emotion control through free-form textual descriptions.
This subtask balances multi-speaker control with basic emotion expression, targeting applications like story narration and gaming NPCs where multiple AI speakers are required.

\textbf{AgentTTS} focuses on generating speech for a single, expressive virtual agent. While it also uses captions like EmoCapTTS, it refines broad emotional categories into more fine-grained states and captures nuanced differences between emotions (e.g., fearful vs. panicked), models interactions between emotional states and low-level speaking styles such as pitch and speed (e.g., happy and slow vs. happy and fast), and incorporates expressive non-speech vocalizations (e.g., sighs, laughter, sobs).
This task closely reflects real-world scenarios in building customized dialogue agents, customer service bots, AI therapists, and other conversational AI applications.

\subsection{Datasets}
\label{datasets}
In this section, we introduce the data collection process in CapSpeech\footnote{\href{https://huggingface.co/datasets/OpenSound/CapSpeech}{https://huggingface.co/datasets/OpenSound/CapSpeech}}
and summarize the data sources used across different tasks in Table~\ref{tab:data2}. The total amount of annotated data is 30.8k hours for age, 33.5k hours for gender, 0.4k hours for timbre, 	33.5 k hours for pitch, 2.5 k hours for accent, 33.5 k hours for speaking rate, 2.4 k hours for emotion, 33.5 k hours for expressiveness of tone, and 2.7 k hours for volume.
\begin{table}[H]
  \caption{CapSpeech data sources used across different tasks. \textit{Italics} indicate machine-labeled data. Regular text indicates human-labeled data. \textcolor{MyLightBlue}{\textbf{Blue}} denotes newly annotated data provided by us, and \text{\musEighth} indicates new audio samples.}
  \label{tab:data2}
  \footnotesize
  \centering
  \begin{tabular}{c|c}
    \toprule
    \textbf{Task} & \textbf{Audio Source} \\
    \midrule
    \multirow{2}{*}{Pretraining} 
      & \textit{Emilia-en}, \textit{\textcolor{MyLightBlue}{\textbf{GigaSpeech}}}, \textit{\textcolor{MyLightBlue}{\textbf{MLS-en}}},  \\
      & \textit{\textcolor{MyLightBlue}{\textbf{CommonVoice}}}, \text{\musEighth} \textit{\textcolor{MyLightBlue}{\textbf{CapSpeech-PT-SEDB}}}\\
    \midrule
    CapTTS, EmoCapTTS, 
      & \textcolor{MyLightBlue}{\textbf{LibriTTS-R}}, EARS, Expresso, \\
      AccCapTTS& \textcolor{MyLightBlue}{\textbf{VCTK}}, VoxCeleb, VoxCeleb2 \\
    \midrule
    CapTTS-SE 
      & \text{\musEighth} \textcolor{MyLightBlue}{\textbf{CapSpeech-SEDB}} \\
    \midrule
    AgentTTS 
      & \text{\musEighth} \textcolor{MyLightBlue}{\textbf{CapSpeech-AgentDB}} \\
    \bottomrule
  \end{tabular}
\end{table}

\subsubsection{Pretraining Sets}
\label{pretraining_set}
CapSpeech includes two pretraining tasks: CapTTS-PT and CapTTS-SE-PT, which are trained jointly during the pretraining phase.

CapTTS-PT comprises four English speech corpora: the English portion of Emilia \cite{DBLP:conf/slt/HeSWLGHLYLSWCZW24}, the English portion of Multilingual LibriSpeech (MLS) \cite{DBLP:conf/interspeech/PratapXSSC20}, GigaSpeech \cite{DBLP:conf/interspeech/ChenCWDZWSPTZJK21}, and CommonVoice \cite{DBLP:conf/lrec/ArdilaBDKMHMSTW20}.
Since the overall audio quality of CommonVoice is poor, we only include the age groups under 20 and over 65, which are not well covered in other datasets.
These datasets cover a wide range of speech sources: LibriVox, YouTube, Podcasts, Audiobooks, and etc.
For Emilia, we directly use the style annotations provided in ParaSpeechCaps \cite{DBLP:journals/corr/abs-2503-04713}, which include 59 diverse style tags.
In total, there are 9,053,734 speech–caption pairs.

We annotate age, gender, pitch, expressiveness of tone, and speaking rate for MLS, GigaSpeech, and CommonVoice, and use an LLM\footnote{\href{https://huggingface.co/mistralai/Mistral-7B-Instruct-v0.3}{https://huggingface.co/mistralai/Mistral-7B-Instruct-v0.3}} to generate natural language captions based on these traits.
Althouth these datasets provide transcriptions, there are some audio files with noisy labels or multiple speakers.
Thus, we apply three data cleaning steps:
(i) samples with durations between 3 and 18 seconds are selected.
(ii) we filter out samples with word error rates (WER) greater than $25\%$, estimated using the OpenAI Whisper toolkit\footnote{\href{https://huggingface.co/openai/whisper-large-v3-turbo}{https://huggingface.co/openai/whisper-large-v3-turbo}} \cite{DBLP:conf/icml/RadfordKXBMS23};
(iii) we remove noisy utterances with estimated signal-to-noise ratios (SNR) below 20 dB, computed using the Squim Objective metric \cite{DBLP:conf/icassp/KumarTNMZHX23}. 
We use a pre-trained age and gender estimator\footnote{\href{https://github.com/audeering/w2v2-age-gender-how-to}{https://github.com/audeering/w2v2-age-gender-how-to}} \cite{burkhardt2023age} to annotate speaker demographics, if they are not provided by the dataset creator.
The age groups are divided into five categories: "child" (1–12 years), "teenager" (13–19 years), "young adult" (20–39 years), "middle-aged adult" (40–64 years), and "elderly" (65 years and older).
Following Parler-TTS \cite{DBLP:journals/corr/abs-2402-01912}, pitch and expressiveness are measured using speaker-level mean and utterance-level standard deviation of pitch, computed with the PENN library\footnote{\href{https://github.com/interactiveaudiolab/penn}{https://github.com/interactiveaudiolab/penn}}.
The speaker-level mean is used to generate a label for speaker pitch relative to gender, and the standard deviation is used as a proxy for how monotone or animated an individual utterance is.
Speaking rate is calculated by dividing the number of phonemes in the transcription by the total duration of the silence-removed utterance. 

Next, we convert all the variables described above into natural language sentences. To do so, we first generate keyword representations for each variable. Discrete labels provided by the dataset creators, such as gender, are used directly as keywords. For continuous variables like pitch and speaking rate, we apply binning to map them into discrete categories. 
After binning all continuous variables, we obtain a complete set of keywords for gender, accent, pitch, expressiveness of tone, and speaking rate.

More specifically, the age category includes "child" ($0.15\%$), "teenager" ($0.20\%$), "young adult" ($32.72\%$), "middle-aged adult" ($63.44\%$), and "elderly" ($3.49\%$); the gender category includes "male" ($57.48\%$) and "female" ($42.52\%$); the pitch category includes "very low-pitch" ($3.22\%$), "low-pitch" ($7.19\%$), "slightly low-pitch" ($26.26\%$), "moderate pitch" ($28.05\%$), "slightly high-pitch" ($21.34\%$), "high-pitch" ($12.45\%$) and "very high-pitch" ($1.49\%$); the expressiveness of tone category includes "very monotone" ($10.59\%$), "monotone" ($48.56\%$), "slightly expressive and animated" ($24.58\%$), "expressive and animated" ($8.32\%$) and "very expressive and animated" ($7.96\%$); the speaking rate category includes "very slowly" ($0.01\%$), "slowly" ($1.60\%$), "slightly slowly" ($15.03\%$), "moderate speed" ($59.81\%$), "slightly fast"($18.98\%$), "fast" ($4.10\%$) and "very fast" ($0.45\%$).

We then apply the LLM to generate natural language captions based on these keywords. To guide the model, we construct a prompt and curate a diverse set of manually verified examples. During caption generation, random examples are paired with different audio files to enhance the diversity of expressions. Finally, we apply post-processing to remove excessively long captions. Below is the LLM prompts used:

\begin{tcolorbox}[colback=white, colframe=black, boxrule=0.4pt,
  title=\textbf{Prompt Design for Pretraining Sets}, fonttitle=\bfseries]
  \textbf{System:} You are a voice style caption generator.\\
  \textbf{User:} Please generate a single text description of a speech sample
  using the provided keywords.\\
  \textbf{Definition of Keywords:}\\
• Gender (e.g., male, female)\\
• Age (e.g., teenager, young adult, etc.)\\
• Tone (e.g., very monotone, quite expressive, etc.)\\
• Pace (e.g., very slowly, quite fast, etc.)\\
• Pitch (e.g., very low pitch, quite high pitch, etc.)\\
\textbf{Instructions}:\\
• Use the keywords to create a grammatically correct and easy-to-understand description of the speech sample, varying the sentence structure and phrasing as much as possible across examples.\\
• Rearrange the keyword order, split ideas across multiple sentences, or introduce descriptive transitions to make the description fluid and natural.\\
• Substitute synonymous terms where appropriate, and rephrase parts of the description to add variety and keep it engaging.\\
• Explore different words to convey the tone, pace, and other characteristics.\\
• You can drop some of the keywords for diversity.\\
• Return only the generated description.\\
Please review the following examples: \textbf{\textit{\textcolor{MyLightBlue}{[randomly selected human-written style captions.]}}}\\
\textbf{Keywords:} 
Gender: \textbf{\textit{\textcolor{MyLightBlue}{[e.g., female]}}};
Age: \textbf{\textit{\textcolor{MyLightBlue}{[e.g., teenager]}}};
Tone: \textbf{\textit{\textcolor{MyLightBlue}{[e.g., bright, expressive]}}};
Pace: \textbf{\textit{\textcolor{MyLightBlue}{[e.g., quick]}}};
Pitch: \textbf{\textit{\textcolor{MyLightBlue}{[e.g., high]}}}.

  \textbf{LLM Assistant:} 
  \textbf{\textit{\textcolor{MyLightPurple}{[e.g., A teenage girl speaks
  with a high-pitched, bright voice that's lively and expressive, delivering her
  words at a quick pace.]}}}
\end{tcolorbox}

To assess potential biases introduced by machine annotation, we conducted a systematic human evaluation on the CapTTS-PT dataset. We randomly sampled 500 examples, each annotated independently by two experienced audio engineers. For each example, individual tags (e.g., age, gender, speaking rate) were evaluated with a binary correctness label (0 or 1) indicating whether the machine-predicted tag accurately reflected the speech. Also, a caption-level quality score was assigned on a 1–5 Likert scale to assess the overall coherence, coverage, and naturalness of the generated captions. Table~\ref{tab:pt_result} shows the detailed results, reporting the mean values along with their $95\%$ confidence intervals. We can see that the average per-tag correctness is $95.5\%$, demonstrating reliable labeling. The average caption quality score is 4.2 out of 5, indicating strong overall fidelity.

\begin{table}[H]
  \caption{Human evaluation on the CapTTS-PT dataset.}
  \label{tab:pt_result}
  \footnotesize
  \centering
  \begin{tabular}{c|c}
    \toprule
    \textbf{Attribute} & \textbf{Score} \\
    \midrule
    Age	&0.979 ± 0.019\\
Gender	&1.000 ± 0.000\\
Pitch	&0.944 ± 0.030\\
Speaking rate	&0.991 ± 0.012\\
Expressive of tone	&0.863 ± 0.004\\
Caption	&4.220 ± 0.071\\
    \bottomrule
  \end{tabular}
\end{table}

For CapTTS-SE-PT, we simulated data using the LibriTTS-R speech corpus and three sound event corpora: VGGSound \cite{DBLP:conf/icassp/ChenXVZ20}, FSDKaggle2018 \cite{DBLP:conf/dcase/FonsecaPFEFPS18}, and ESC-50 \cite{piczak2015dataset}. 
We curate a collection of 394 distinct sound events drawn from the categories of these datasets.
We refer to this dataset as \textbf{CapSpeech-PT-SEDB}.
Sound events are introduced into speech using two modes: (i) insertion, where the event is inserted at a specific location; and (ii) background, where the event is layered beneath the speech.
For the background mode, we use noisy audio from VGGSound, while the insertion mode relies on cleaner sources from FSDKaggle2018 and ESC-50. For each LibriTTS-R sample, a sound event and mode are randomly selected. Forced alignment is performed using the Montreal Forced Aligner\footnote{\href{https://huggingface.co/datasets/cdminix/libritts-r-aligned}{https://huggingface.co/datasets/cdminix/libritts-r-aligned}} to determine insertion points—either at the beginning, middle, or end of the utterance.
To ensure the simulated audio sounds natural and smooth, we select insertion points between words with a minimum interval of 0.3 seconds, ensuring a pause that does not disrupt word continuity.
The speech and sound event are then mixed at a random SNR ranging from –3 dB to 6 dB. To scale up the data, each speech sample is simulated five times with different configurations.

\begin{table}[h]
  \caption{Sound event examples in CapSpeech.}
  \label{tab:events}
  \scriptsize
  \centering
    \begin{tabular}{llll}
    \toprule
         Applause& Bark& Burping or eructation& Bus\\
         Chime&
        Writing& Drawer open or close& Cough\\ Fart&
        Finger snapping&
        Fireworks&
        Crow\\
        Keys jangling& Knock& Laughter&
        Meow\\ Microwave oven&Clock tick&
        Shatter&	Hand saw\\ Squeak& Tearing& Telephone& Computer\\ keyboard
        &Dog&Rooster&	Rain\\	Crying baby&	Door knock&	Scissors&Helicopter\\	Sea waves&	Sneezing&	Mouse click&	Chainsaw\\
        Pig&	Crackling fire&	Clapping&	Keyboard typing\\	Siren&
        Cow&	Crickets&	Breathing\\	Door wood creaks&	Car horn&
        Frog&	Chirping birds\\	Coughing&	Can opening&	Engine&
        Cat\\	Water drops&	Footsteps&	Washing machine & Hand saw\\	Train&
        Hen&	Wind&	Laughing\\	Vacuum cleaner&	Church bells&
        Insects (flying)&	Pouring water\\	Brushing teeth&	Clock alarm&	Airplane&
        Sheep\\	Toilet flush&	Snoring&	Gunshot or gunfire&	Thunderstorm	\\		Glass &breaking&Drinking, sipping\\
    \bottomrule
    \end{tabular}
\end{table}

In total, the dataset contains 10,054,530 speech–caption pairs, with 7,894 pairs reserved for validation and 7,959 pairs held out for testing.

\subsubsection{SFT Sets for CapTTS, EmoCapTTS, and AccCapTTS}
We group CapTTS, EmoCapTTS, and AccCapTTS together because they share both underlying speech corpora and common stylistic attributes (e.g., timbre, speaking rate). To support these tasks, we aggregate human-annotated data from six publicly available corpora: LibriTTS-R \cite{DBLP:conf/interspeech/KoizumiZKDYMB0H23}, VCTK \cite{Yamagishi2019VCTK}, VoxCeleb \cite{DBLP:conf/interspeech/NagraniCZ17}, VoxCeleb2 \cite{DBLP:conf/interspeech/ChungNZ18}, EARS \cite{DBLP:journals/corr/abs-2406-06185}, and Expresso \cite{DBLP:conf/interspeech/NguyenHDSGFRCSH23}. The aggregation process involves directly using existing captions, enhancing captions with additional speaker traits, and generating captions using a large language model based on structured labels. For LibriTTS-R, we incorporate annotations from LibriTTS-P \cite{DBLP:journals/corr/abs-2406-07969} and DreamVoiceDB \cite{hai2024dreamvoice}, which particularly focus on intrinsic speaker traits, and use the LLM to generate captions. The prompt used for generating LibriTTS-P annotations is shown below.

\begin{tcolorbox}[colback=white, colframe=black, boxrule=0.4pt,
  title=\textbf{Prompt Design for LibriTTS-P}, fonttitle=\bfseries]
  \textbf{System:} You are a voice style caption generator.\\
  \textbf{User:} \\
  • Given a list of keywords related to a speaker's vocal traits, generate 10 diverse and natural-sounding style captions.\\
• You can freely drop or combine keywords. Vary the sentence structures and expressions as much as possible.\\
• Respond with only the caption.\\
\textbf{Keywords:} \textbf{\textit{\textcolor{MyLightBlue}{[e.g., teenager, female, bright, smooth, nasal, cute, and quick pace.]}}}\\
  \textbf{LLM Assistant:} \textbf{\textit{\textcolor{MyLightPurple}{[e.g., A teenage girl’s voice is bright and smooth, with a slight nasal quality, and she speaks at a lively, quick pace.]}}}
\end{tcolorbox}

From DreamVoiceDB, we also use the VCTK portion, which includes accent annotations, and use the LLM to generate captions. The prompt used for this caption generation is shown below.
\begin{tcolorbox}[colback=white, colframe=black, boxrule=0.4pt,
  title=\textbf{Prompt Design for VCTK}, fonttitle=\bfseries]
  \textbf{System:} You are a voice style caption generator.\\
  \textbf{User:} \\
  • Please naturally incorporate the accent information into the following description, without changing its original meaning. \textbf{Accent:} \textbf{\textit{\textcolor{MyLightBlue}{[e.g., Indian Accent.]}}} \textbf{Original description:} \textbf{\textit{\textcolor{MyLightBlue}{[e.g., A senior woman's voice carries with warmth, depth, and an authoritative tone.]}}} \\
• You can freely vary the sentence structures and expressions as much as possible.\\
• Respond with only the modified caption.\\
  \textbf{LLM Assistant:} \textbf{\textit{\textcolor{MyLightPurple}{[e.g., A senior woman’s voice resonates with warmth and depth, layered with an authoritative tone and gently marked by an Indian accent.]}}}
\end{tcolorbox}

For VoxCeleb, VoxCeleb2, EARS, and Expresso, we apply annotations from ParaSpeechCaps \cite{DBLP:journals/corr/abs-2503-04713}, which cover both intrinsic and expressive style traits.

Based on the collected and annotated captions, we construct SFT datasets to support three tasks:
(1) CapTTS serves as a general-purpose task for caption-based speech synthesis. It leverages all above sources, resulting in 347,783 speech–caption pairs, with 18,348 used for validation and 20,756 held out for testing.
(2) EmoCapTTS focuses on emotional expression, using data from the EARS and Expresso corpora, which offer high-quality emotion annotations. This subset contains 26,428 samples, with 1,800 used for validation and 1,937 reserved for testing.
(3) AccCapTTS targets accent control, utilizing data from VCTK, VoxCeleb, and VoxCeleb2, which offer reliable accent annotations. It comprises 113,197 samples, with 10,599 used for validation and 13,051 held out for testing.

\subsubsection{SFT Set for CapTTS-SE}
To the best of our knowledge, there are \textbf{no existing open-source} TTS datasets that include sound events. To address this gap, we propose a new human-processed high-quality dataset for the CapTTS-SE task, named \textbf{CapSpeech-SEDB}. CapSpeech-SEDB comprises 500 audio mixtures incorporating 10 common sound events (\textit{i.e.} the sound of coughing, laughing, clapping, can opening, footsteps, keyboard typing, alarm clock, door knocking, dog barking, and cat meow), meticulously crafted by \textbf{five audio engineers} with expertise in music production or film sound design. 
To ensure the simulated audio sounds natural and smooth, we select insertion and background points between words with a minimum interval of 0.3 seconds, ensuring a pause that does not disrupt word continuity. 
For each speech recording and its corresponding set of sound event candidates, the engineers select the most appropriate event and follow detailed instructions—such as inserting the event or using it as background—to produce a new mixture. They fine-tune the result by balancing volume levels, applying fade-ins and fade-outs, and adjusting the equalizer and compressor to achieve a natural and seamless integration of speech and sound events. Moreover, we create another 500 transcription-caption pairs for testing.

\subsubsection{SFT Set for AgentTTS}
To the best of our knowledge, there are \textbf{no existing open-source} datasets that feature single-speaker recordings paired with diverse and fine-grained emotion and style prompts, suitable for applications such as customer service and AI therapy, which require nuanced expressiveness and broad stylistic control.
To address this gap, we present \textbf{CapSpeech-AgentDB}, a new dataset comprising 10{,}000 caption-speech pairs, totaling approximately 25.2 hours of high-quality recordings by a single female speaker, in which 500 pairs are held out for testing. The dataset captures subtle gradations between emotional states (e.g., disappointed vs. sadness vs. grief, annoyed vs. angry vs. irate), and includes less commonly represented emotions (e.g., curious, jealous, resentful, focused, distracted), which are rarely present in existing speech emotion corpora. In addition, it features a wide variety of speaking styles (e.g., laughing, crying, panting, whispering), as well as nuanced combinations of emotions and styles (e.g., angry growl vs. angry scream, happy and slow vs. happy and fast), enabling fine-grained control over both expressive and stylistic variation.

To collect the recordings, we first generate prompt-content pairs, paying careful attention to balancing subcategories of emotions and styles. The content spans a wide range of conversational scenarios, including everyday dialogue, cafés, hospitals, classrooms, sci-fi films, horror movies, and more. During recording, the speaker follows the prompts to perform the specified emotion and style, and is allowed to revise the prompt or content if any part is unclear or unnatural.

The following prompt template guides an LLM in generating expressive prosody descriptions and corresponding dialogue samples. As this task is more complex, we use GPT-4o-mini\footnote{\href{https://platform.openai.com/docs/models/gpt-4o-mini}{https://platform.openai.com/docs/models/gpt-4o-mini}} for better quality and naturalness. Below is the prompts used in this process.

\begin{tcolorbox}[colback=white, colframe=black, boxrule=0.4pt,
  title=\textbf{Prompt Design for CapSpeech-AgentDB}, fonttitle=\bfseries]
  \textbf{System:} You are a speech and voice expert contributing to an expressive speech dataset.\\
  \textbf{User:} Please generate a natural-language prosody prompt that vividly describes how the speaker’s voice would sound, given a high-level expression (e.g., emotion, mental state) and a low-level expression (e.g., pitch, pace, volume). \\
  \textbf{Guidelines:} \\
• Ensure that the prosody description is vivid, coherent, and captures the expression naturally. \\
• It is acceptable to merge or omit specific low-level features for fluency. \\
• Please follow these examples: \textbf{\textcolor{MyLightBlue}{\textit{[random sample 3 human-written style-captions]}}}\\
\textbf{LLM Assistant:} \textbf{\textit{\textcolor{MyLightPurple}{[e.g., The voice is quick and unsteady, occasionally faltering, revealing anxiousness.]}}}\\
\textbf{User:} 
Using the above prosody prompt and the provided conversation scene, generate a single line of dialogue or a short dialogue excerpt from only one speaker.

\textbf{Guidelines:} \\
• Ensure the speech sounds natural and is emotionally aligned with the described vocal tone and context. \\
• Use varied sentence structures and lexical choices across examples to enhance dataset diversity. \\
• Do not include or imply any lines from other characters—generate content exclusively for one speaker.\\
• \textcolor{gray}{Optional: When appropriate, incorporate non-speech vocalizations from the provided list ([Gasp], [Laughter], ..., [Yawn]) to enhance the expressiveness of the speech.}

\textbf{Conversation scene:} \textbf{\textcolor{MyLightBlue}{\textit{[e.g., in a horror movie]}}}\\
\textbf{LLM Assistant:} 
\textbf{\textit{\textcolor{MyLightPurple}{[e.g., 
I—I think I saw something move behind the curtain. No, no, I’m not imagining it. It was real. I swear it was real. We can’t stay here. We have to go—now!]}}}
\end{tcolorbox}

These outputs form the basis of the CapSpeech-AgentDB, providing reference material for voice actors to produce the final audio recordings. The set of high-level emotional expressions, low-level prosody expressions, and non-speech vocalizations used to guide the prompts is presented in Table~\ref{tab:expressive_voice_labels}.

\begin{table*}[t]
  \caption{List of Expressive Voice Labels}
  \label{tab:expressive_voice_labels}
  \centering
  \scriptsize
  \begin{tabular}{l|lllllll}
    \toprule
    \multirow{7}{*}{\textbf{High-level Emotional Expressions}}& 
    Neutral & Aggressive & Angry & Annoyed & Contemptuous & Hateful & Raging  \\
    & Disgusted & Sarcastic & Resentful & Alarmed & Apprehensive & Fearful & Nervous \\
    & Panicked & Terrified& Accepting & Admiring & Amazed & Excited & Hopeful   \\
    & Interested & Joyful & Loving& Peaceful & Surprised & Disappointed & Dismayed \\
    & Distressed & Grieving  & Hopeless & Sad  &Authoritative & Respectful & Sympathetic \\
     & Trusting & Anticipating & Curious & Eager & Focused& 
    Tired & Awkward \\
    & Confused & Distracted & Skeptical & Bored & Envious & Submissive & Shameful\\
    \midrule
    \multirow{2}{*}{\textbf{Low-level Prosody Expressions}}& Default & Monotonous & Deep & Sharp & Gentle & Loud & Mumble \\ & Stutter
    Whispering & Crying & Laughing & & & \\
    \midrule
    \multirow{2}{*}{\textbf{Non-speech Voices}}&Aww & Throat-clearing & Cheering & Contemplation & Gasp & Groan & Laughter\\
    & Panting
    Scream & Sigh & Sneering laughter & Sob & Yawn & \\
    \bottomrule
  \end{tabular}
\end{table*}


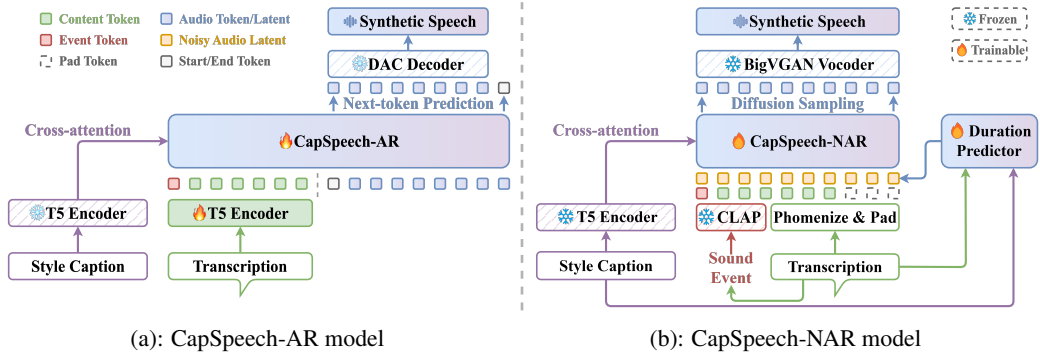
\begin{figure*}[t]
  \centering

  \begin{subfigure}[t]{0.48\linewidth}
    \centering
    \includegraphics[width=\linewidth]{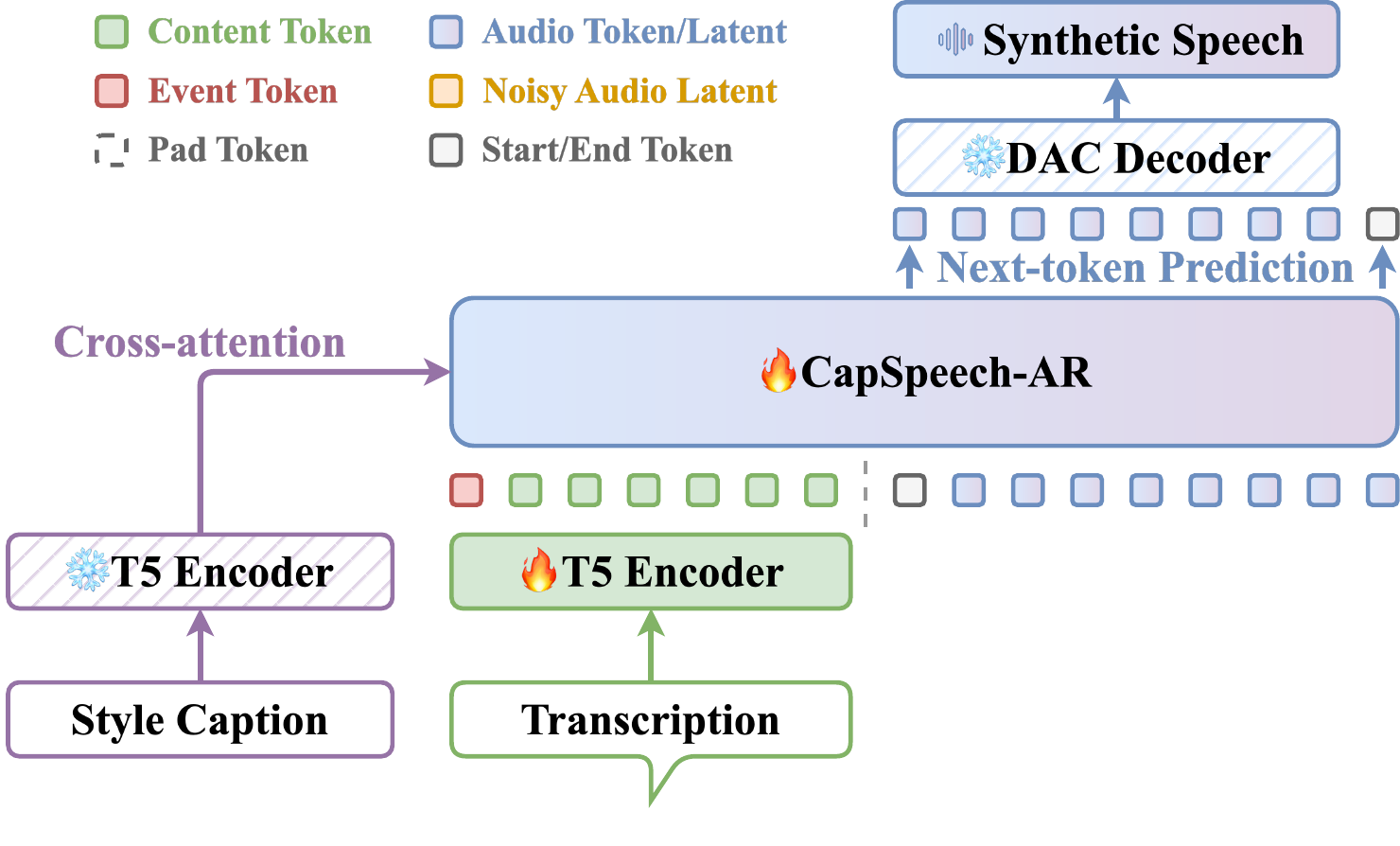}
    \caption{CapSpeech-AR model}
    \label{fig:subfig1}
  \end{subfigure}
  %
  \begin{minipage}[t]{0.01\linewidth}
    \centering
    \begin{tikzpicture}
      \draw[dashed, line width=1pt, color=gray!60] (0,0) -- (0, 4.1);
    \end{tikzpicture}
  \end{minipage}
   \begin{subfigure}[t]{0.48\linewidth}
    \centering
    \includegraphics[width=\linewidth]{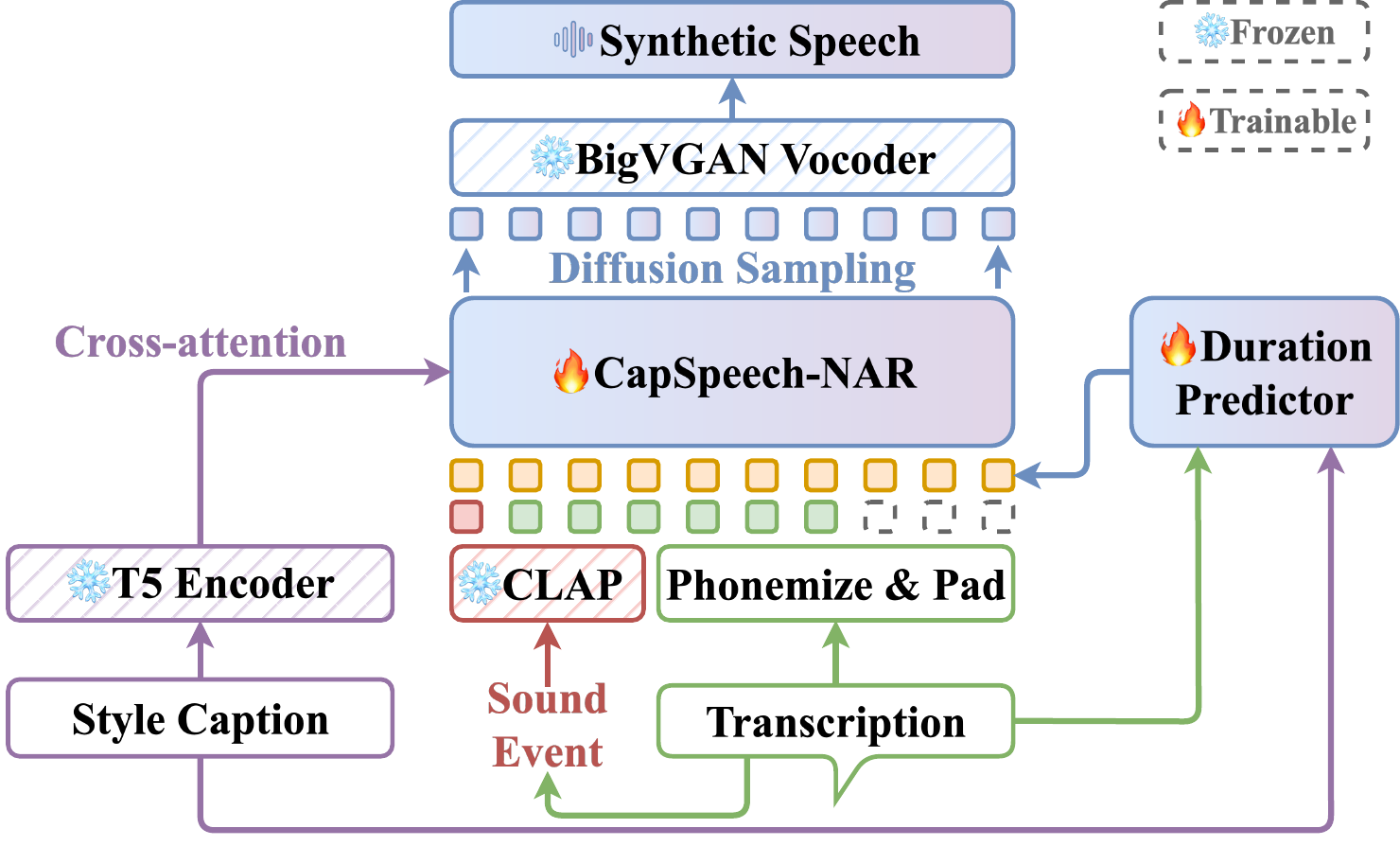}
    \caption{CapSpeech-NAR model}
    \label{fig:subfig2}
  \end{subfigure}

  \caption{Illustration of CapSpeech model architectures.}
  \label{fig:model}
\end{figure*}


\section{Experiments}
\subsection{Baselines}
\label{baselines}
We adopt two strong baselines for the CapSpeech benchmark.

\textbf{CapSpeech-AR}: The first is Parler-TTS\footnote{\href{https://github.com/huggingface/parler-tts}{https://github.com/huggingface/parler-tts}} \cite{DBLP:journals/corr/abs-2402-01912}, a state-of-the-art AR method based on a codec language model.
We use the 44.1kHz version of Descript Audio Codec (DAC) \cite{DBLP:conf/nips/KumarSLKK23} to provide discrete audio representations. A delay pattern is applied to handle multiple codebooks \cite{DBLP:conf/nips/CopetKGRKSAD23}, and cross-attention is used to incorporate caption-based style conditioning.
FLAN-T5\footnote{\href{https://huggingface.co/google/flan-t5-large}{https://huggingface.co/google/flan-t5-large}} \cite{DBLP:journals/jmlr/ChungHLZTFL00BW24} is used to extract the text features from the transcription and style caption.
We retain the original architecture, with the exception of adding special tokens to support the CapTTS-SE task. The FLAN-T5 model used in the transcription is fine-tuned.
As illustrated in Fig.~\ref{fig:tasks}, transcriptions in CapTTS-SE support two modes for integrating sound events: (i) \textbf{Background mode} and (ii) \textbf{Insertion mode}.
sound event tokens (e.g., \texttt{<telephone>}, \texttt{<knock>}) are placed at the beginning of the sequence. The tags \texttt{<B>} and \texttt{</B>} mark the start and end of a background sound segment, while \texttt{<I> </I>} denote the insertion point.
This design allows flexible control over the placement and type of sound events within the synthesized speech.

\textbf{CapSpeech-NAR}: In addition, we adopt F5-TTS\footnote{\href{https://github.com/SWivid/F5-TTS}{https://github.com/SWivid/F5-TTS}} \cite{DBLP:journals/corr/abs-2410-06885}, a state-of-the-art NAR method based on flow matching with a Diffusion Transformer.
In our adaptation, we remove the audio-prompt masking component and instead use cross-attention to incorporate caption-based style conditioning.
BigVGAN\footnote{\href{https://huggingface.co/nvidia/bigvgan_v2_24khz_100band_256x}{https://huggingface.co/nvidia/bigvgan\_v2\_24khz\_100band\_256x}} \cite{DBLP:conf/iclr/LeePGCY23} is used as the vocoder, and we apply QK-Norm \cite{DBLP:conf/emnlp/HenryDPC20} to stabilize training. Transcriptions are processed using grapheme-to-phoneme conversion (g2p)\footnote{\href{https://github.com/Kyubyong/g2p}{https://github.com/Kyubyong/g2p}}, and we insert special tokens \texttt{<B>}, \texttt{</B>}, \texttt{<I>}, and \texttt{</I>} to indicate background and insertion-based sound events.
To enhance generalization, we do not include sound event tags directly in the input sequence. Instead, we extract LAION-CLAP
\cite{DBLP:conf/icassp/WuCZHBD23} embeddings for the specified sound event and use them as the additional input.
This design allows the model to generalize to unseen sound events during inference.
Similar to the AR model, a Flan-T5 model is used to extract the text features from the style caption.
Since the NAR model cannot predict audio duration directly, inspired by \cite{DBLP:journals/corr/abs-2408-13893}, we fine-tune a BERT model\footnote{\href{https://huggingface.co/google-bert/bert-base-uncased}{https://huggingface.co/google-bert/bert-base-uncased}} \cite{DBLP:journals/corr/abs-1810-04805} that takes both the transcription and the caption as input to estimate the total duration of the entire audio.

\subsection{Implementation Details}
\label{settings}

Table~\ref{tab:model} summarizes the computation cost and model size of AR and NAR models. In the Transformer column, the numbers denote the model dimension, the number of layers, the number of heads, and the multiples of hidden size. Training time is calculated on 8 NVIDIA-H100 GPUs and Real-time-factor (RTF) is calculated on a single NVIDIA-A100 GPU.

\begin{table}[H]
  \caption{Model complexity and computational cost.}
  \label{tab:model}
  \footnotesize
  \centering
  \begin{tabular}{c|cccc}
    \toprule
    Model& Transformer & Params(M) & Training Time & RTF  \\
    \midrule
    CapSpeech-AR &1024,24,16,4&880&24 Days& 1.24\\
    CapSpeech-NAR &1024,24,16,4&724&10 Days& 1.01\\
    \bottomrule
  \end{tabular}
\end{table}

During pretraining of the AR model, we use a batch size of 32, a learning rate of 1e-3, a weight decay of 0.01, and gradient accumulation over 6 steps. Linear warm-up is applied for the first 5000 steps. The model is trained for 10 epochs, taking approximately 24 days on 8 NVIDIA H100 GPUs.
For fine-tuning the AR model, the learning rate is set to 1e-4. Warm-up is applied for 1000 steps for the CapTTS, EmoCapTTS, and AccCapTTS tasks, and for 100 steps for the AgentTTS and CapTTS-SE tasks. Fine-tuning on CapTTS, EmoCapTTS, and AccCapTTS takes 47 hours, 24 hours, and 24 hours, respectively (over 5 epochs) on a single NVIDIA A100 GPU. Fine-tuning on AgentTTS and CapTTS-SE takes 14 hours and 1 hour, respectively (over 50 epochs) on a single NVIDIA A100 GPU.

For the NAR model, pretraining is conducted with a batch size of 512, a learning rate of 2e-4, and a weight decay of 0.01. A linear warm-up of 5000 steps is followed by linear decay over the remaining training steps. The model is trained for 10 epochs, requiring approximately 10 days on 8 NVIDIA H100 GPUs.
During fine-tuning of the NAR model, the learning rate is set to 2e-5. Warm-up is applied for 1000 steps for CapTTS, EmoCapTTS, and AccCapTTS, and for 100 steps for AgentTTS and CapTTS-SE. Fine-tuning on CapTTS, EmoCapTTS, and AccCapTTS takes 69 hours, 33 hours, and 33 hours, respectively (over 15 epochs) on a single NVIDIA A100 GPU. Fine-tuning on AgentTTS and CapTTS-SE takes 6.8 hours and 0.6 hours, respectively (over 50 epochs) on a single NVIDIA A100 GPU.
For the duration predictor, we fine-tune the \textit{bert-base-uncased} model with a learning rate of 1e-4 and a batch size of 256. The model is trained on the CapSpeech pretraining set for 2 epochs using a single NVIDIA A100 GPU, which takes approximately 70 hours. This predictor is used for the CapTTS, CapTTS-SE, EmoCapTTS, and AccCapTTS tasks. For the AgentTTS task, we further fine-tune the model for the single female speaker using a learning rate of 1e-5 for 10 epochs, which takes about half an hour.

All models are trained using the AdamW optimizer \cite{DBLP:conf/iclr/LoshchilovH19}.
Following \cite{DBLP:journals/corr/abs-2503-04713}, we initialize the AR model with the \textit{Parler-TTS Mini v1}\footnote{\href{https://huggingface.co/parler-tts/parler-tts-mini-v1}{https://huggingface.co/parler-tts/parler-tts-mini-v1}} model. During inference, we perform AR generation with a temperature of 1.0 and a repetition penalty of 1.0, and introduce inference-only classifier-free guidance (CFG) \cite{DBLP:journals/corr/abs-2409-07556}. For the NAR model, following \cite{DBLP:journals/corr/abs-2410-06885}, we use the default CFG strength of 2 and a Sway Sampling coefficient of $-1$. 
To avoid any potential misuse, we introduce a novel watermarking technology \cite{DBLP:journals/corr/abs-2409-07556} to each model, and open-source our
code and model weights under CC BY-NC 4.0 license.

\subsection{Metrics}
\textbf{Objective Metrics:}
We evaluate three key aspects of model performance: style consistency, audio quality, and intelligibility.
For style consistency, we compute classification accuracy across multiple categories: age, gender, pitch, expressiveness of tone, speed, accent, and emotion. The average accuracy across these attributes is reported as \textbf{Style-ACC}. 
Following Section~\ref{pretraining_set}, we utilized existing toolkits to predict various speaker attributes, including age, gender, pitch, vocal expressiveness, and speaking rate. 
For emotion and accent classification, we introduced the state-of-the-art classifiers in \cite{feng2025vox}.
The emotion classifier includes nine emotion labels: Anger, Contempt, Disgust, Fear, Happiness, Neutral, Sadness, Surprise, and Other. The accent classifier covers 16 accent categories: East Asia, English, Germanic (e.g., German), Irish, North America, Northern Irish, Oceania (e.g., Australia), Romance (e.g., Spanish, French), Scottish, Semitic, Slavic (e.g., Russian, Polish), South African, Southeast Asia, South Asia, Welsh, and Other accents not covered above.

Audio quality is assessed using UTMOSv2\footnote{\href{https://github.com/sarulab-speech/UTMOSv2}{https://github.com/sarulab-speech/UTMOSv2}} \cite{DBLP:conf/slt/BabaNSS24}.
For intelligibility, we compute a text-normalized WER between the ASR transcript of the generated speech and the input transcript. This is done using \textit{openai/whisper-large-v3-turbo} along with the Whisper text normalizer.

\textbf{Subjective Metrics:}
We experimented with state-of-the-art speech understanding models
\cite{DBLP:journals/corr/abs-2407-10759,kimiteam2025kimiaudiotechnicalreport} to evaluate style consistency, but they failed to generate high-quality speech captions.
Following \cite{DBLP:journals/corr/abs-2503-04713}, we recruit 15 native speakers via Prolific\footnote{\href{https://www.prolific.com/}{https://www.prolific.com/}} to evaluate three subjective aspects: Style Consistency MOS (SMOS), Naturalness MOS (NMOS), and Intelligibility MOS (IMOS).
We recruit high-quality participants from Prolific with a minimum approval rate of 95$\%$, at least 100 prior approved submissions.
Before annotations, we perform a qualification task using 10 manually-selected samples, and select 15 annotators that succeeded on at least 9.
For each task, we randomly select 100 test samples from the test sets and ensure that each sample is rated by three annotators.
For the CapTTS task, annotators are instructed to evaluate overall style consistency. For the EmoCapTTS and AccCapTTS tasks, they are asked to place greater emphasis on emotion and accent characteristics. In the CapTTS-SE task, annotators are instructed to factor in the accuracy of sound events when rating intelligibility. For the AgentTTS task, speaker similarity to the reference recording is also considered as part of the intelligibility score.
Each example is rated by three annotators, and we report the mean scores along with $95\%$ confidence intervals.

\subsection{Results and Analysis}

\begin{table*}[t]
  \caption{Results on the pretraining task.}
  \label{tab:result1}
  \centering
  \begin{tabular}{l|l|ccc|ccc}
    \toprule
    \multirow{2}{*}{Model} & \multirow{2}{*}{Data} & \multicolumn{3}{c|}{Objective Evaluations} & \multicolumn{3}{c}{Subjective Evaluations} \\
    && Style-ACC$\uparrow$ & UTMOSv2$\uparrow$ & WER$\downarrow$ & SMOS$\uparrow$ & NMOS$\uparrow$ & IMOS$\uparrow$ \\
    \midrule
    \multirow{2}{*}{AR}&ParaSpeechCaps&$44.0\%$&$2.93$&$10.8\%$&$3.48 {\scriptstyle \pm 0.12}$&$3.21 {\scriptstyle \pm 0.08}$&$3.88 {\scriptstyle \pm 0.12}$\\
    &\textbf{Ours}&$\boldsymbol{52.2\%}$&$\boldsymbol{3.18}$&$\boldsymbol{9.1\%}$&$\boldsymbol{3.62} {\scriptstyle \pm 0.11}$&$\boldsymbol{3.65} {\scriptstyle \pm 0.10}$&$\boldsymbol{4.23} {\scriptstyle \pm 0.12}$\\
    \midrule
    \multirow{2}{*}{NAR}&ParaSpeechCaps&$51.8\%$&$3.16$&$9.5\%$&$3.55 {\scriptstyle \pm 0.11}$&$3.38 {\scriptstyle \pm 0.09}$&$3.93 {\scriptstyle \pm 0.10}$\\
    &\textbf{Ours}&$\boldsymbol{62.1\%}$&$\boldsymbol{3.43}$&$\boldsymbol{8.8\%}$&$\boldsymbol{3.80} {\scriptstyle \pm 0.13}$&$\boldsymbol{3.79} {\scriptstyle \pm 0.10}$&$\boldsymbol{4.40} {\scriptstyle \pm 0.09}$\\
    \bottomrule
  \end{tabular}
\end{table*}

\begin{table*}[t]
  \caption{Results on the fine-tuning tasks. \textcolor{gray}{$\star$ denotes models without pretraining.} CapTTS-SE results without pretraining are not reported as there are only 500 samples in CapTTS-SEDB. Style-ACC is not applicable to AgentTTS and CapTTS-SE, and UTMOSv2 is not applicable to CapTTS-SE.}
  \label{tab:result4}
  \centering
  \begin{tabular}{l|l|ccc|ccc}
    \toprule
    \multirow{2}{*}{Task} &
    \multirow{2}{*}{Model} &
    \multicolumn{3}{c|}{Objective Evaluations} & \multicolumn{3}{c}{Subjective Evaluations} \\
    & & Style-ACC$\uparrow$ & UTMOSv2$\uparrow$ & WER$\downarrow$ & SMOS$\uparrow$ & NMOS$\uparrow$ & IMOS$\uparrow$ \\
    \midrule
    \multirow{4}{*}{CapTTS}
    & \textcolor{gray}{AR$\star$} & $\textcolor{gray}{42.1\%}$ & $\textcolor{gray}{2.45}$ & $\textcolor{gray}{21.5\%}$ & \textcolor{gray}{$3.21 \scriptstyle \pm 0.17$} & \textcolor{gray}{$3.40 \scriptstyle \pm 0.14$} & \textcolor{gray}{$3.86 \scriptstyle \pm 0.12$} \\
    & AR & $56.0\%$ & $3.02$ & $11.2\%$ & $3.72 \scriptstyle \pm 0.12$ & $3.62 \scriptstyle \pm 0.11$ & $4.15 \scriptstyle \pm 0.10$ \\
    & \textcolor{gray}{NAR$\star$} & $\textcolor{gray}{46.4\%}$ & $\textcolor{gray}{2.61}$ & $\textcolor{gray}{19.5\%}$ & \textcolor{gray}{$3.40 \scriptstyle \pm 0.12$} & \textcolor{gray}{$3.60 \scriptstyle \pm 0.13$} & \textcolor{gray}{$3.95 \scriptstyle \pm 0.10$} \\
    & NAR & $\boldsymbol{66.0\%}$ & $\boldsymbol{3.37}$ & $\boldsymbol{9.2\%}$ & $\boldsymbol{3.85 \scriptstyle \pm 0.13}$ & $\boldsymbol{3.95 \scriptstyle \pm 0.12}$ & $\boldsymbol{4.34 \scriptstyle \pm 0.11}$ \\
    \midrule
    \multirow{2}{*}{CapTTS-SE} & AR & / & / & $7.7\%$ & $3.69 \scriptstyle \pm 0.12$ & $3.52 \scriptstyle \pm 0.14$ & $\boldsymbol{3.45 \scriptstyle \pm 0.14}$ \\
    & NAR & / & / & $\boldsymbol{3.0\%}$ & $\boldsymbol{3.75 \scriptstyle \pm 0.13}$ & $\boldsymbol{3.60 \scriptstyle \pm 0.11}$ & $3.33 \scriptstyle \pm 0.15$ \\
    \midrule
    \multirow{4}{*}{EmoCapTTS}
& \textcolor{gray}{AR$\star$} & $\textcolor{gray}{40.5\%}$ & $\textcolor{gray}{1.98}$ & $\textcolor{gray}{54.8\%}$ & \textcolor{gray}{$2.52 \scriptstyle \pm 0.13$} & \textcolor{gray}{$2.83 \scriptstyle \pm 0.13$} & \textcolor{gray}{$3.05 \scriptstyle \pm 0.14$} \\
& AR & $58.6\%$ & $3.04$ & $11.5\%$ & $3.72 \scriptstyle \pm 0.14$ & $3.58 \scriptstyle \pm 0.12$ & $4.12 \scriptstyle \pm 0.13$ \\
& \textcolor{gray}{NAR$\star$} & $\textcolor{gray}{44.5\%}$ & $\textcolor{gray}{2.10}$ & $\textcolor{gray}{45.5\%}$ & \textcolor{gray}{$2.72 \scriptstyle \pm 0.14$} & \textcolor{gray}{$3.00 \scriptstyle \pm 0.12$} & \textcolor{gray}{$3.28 \scriptstyle \pm 0.12$} \\
& NAR & $\boldsymbol{67.2\%}$ & $\boldsymbol{3.34}$ & $\boldsymbol{10.1\%}$ & $\boldsymbol{3.77 \scriptstyle \pm 0.12}$ & $\boldsymbol{3.88 \scriptstyle \pm 0.13}$ & $\boldsymbol{4.35 \scriptstyle \pm 0.10}$ \\
    \midrule
    \multirow{4}{*}{AccCapTTS}
    & \textcolor{gray}{AR$\star$} & $\textcolor{gray}{37.2\%}$ & $\textcolor{gray}{2.06}$ & $\textcolor{gray}{48.0\%}$ & \textcolor{gray}{$2.98 \scriptstyle \pm 0.10$} & \textcolor{gray}{$3.30 \scriptstyle \pm 0.12$} & \textcolor{gray}{$3.75 \scriptstyle \pm 0.14$} \\
    & AR & $54.9\%$ & $3.10$ & $10.9\%$ & $3.77 \scriptstyle \pm 0.12$ & $3.67 \scriptstyle \pm 0.11$ & $4.20 \scriptstyle \pm 0.11$ \\
    & \textcolor{gray}{NAR$\star$} & $\textcolor{gray}{39.2\%}$ & $\textcolor{gray}{2.41}$ & $\textcolor{gray}{30.2\%}$ & \textcolor{gray}{$3.12 \scriptstyle \pm 0.12$} & \textcolor{gray}{$3.36 \scriptstyle \pm 0.11$} & \textcolor{gray}{$3.88 \scriptstyle \pm 0.12$} \\
    & NAR & $\boldsymbol{66.4\%}$ & $\boldsymbol{3.45}$ & $\boldsymbol{8.8\%}$ & $\boldsymbol{3.91 \scriptstyle \pm 0.12}$ & $\boldsymbol{3.90 \scriptstyle \pm 0.12}$ & $\boldsymbol{4.45 \scriptstyle \pm 0.09}$ \\
    \midrule
    \multirow{4}{*}{AgentTTS}
    & \textcolor{gray}{AR$\star$} & / & $\textcolor{gray}{1.85}$ & $\textcolor{gray}{56.9\%}$ & \textcolor{gray}{$1.50 \scriptstyle \pm 0.22$} & \textcolor{gray}{$2.35 \scriptstyle \pm 0.20$} & \textcolor{gray}{$2.52 \scriptstyle \pm 0.13$} \\
    & AR & / & $\boldsymbol{3.26}$ & $10.2\%$ & $\boldsymbol{3.50 \scriptstyle \pm 0.14}$ & $3.70 \scriptstyle \pm 0.11$ & $4.22 \scriptstyle \pm 0.12$ \\
    & \textcolor{gray}{NAR$\star$} & / & $\textcolor{gray}{1.92}$ & $\textcolor{gray}{54.8\%}$ & \textcolor{gray}{$1.82 \scriptstyle \pm 0.18$} & \textcolor{gray}{$2.10 \scriptstyle \pm 0.15$} & \textcolor{gray}{$2.77 \scriptstyle \pm 0.15$} \\
    & NAR & / & $3.07$ & $\boldsymbol{9.5\%}$ & $3.42 \scriptstyle \pm 0.14$ & $\boldsymbol{3.80 \scriptstyle \pm 0.11}$ & $\boldsymbol{4.41 \scriptstyle \pm 0.10}$ \\
    \bottomrule
  \end{tabular}
\end{table*}

\textbf{Pretraining Stage:} We compare the CapTTS task trained on our proposed CapTTS pretraining set against training on the previous large-scale dataset ParaSpeechCaps (PSC-Scaled partition). Table~\ref{tab:result1} presents the results on the CapTTS task, comparing models using both objective and subjective metrics. 
Models trained on the CapTTS Pretraining set achieve significantly better style consistency, naturalness, and intelligibility than those trained on ParaSpeechCaps, demonstrating the effectiveness of our dataset. Compared to AR models, NAR models consistently achieve better performance across all metrics, highlighting their advancement on the CapTTS task. 

\textbf{Fine-tuning Stage:} We pre-train the AR and NAR models on our proposed CapTTS and CapTTS-SE pretraining sets, and then fine-tune them on the downstream tasks. 
The results, shown in Table~\ref{tab:result4}, demonstrate that pretraining provides substantial benefits across all downstream tasks--particularly for CapTTS-SE and AgentTTS, where training data is limited. Notably, our benchmark indicates that strong style consistency, naturalness, and intelligibility can be achieved on the CapTTS, EmoCapTTS, and AccCapTTS tasks, with the NAR model reaching SMOS, NMOS, and IMOS scores of at least 3.77, 3.88, and 4.34, respectively. Additionally, the AR model surpasses the NAR model on certain metrics in the CapTTS-SE and AgentTTS tasks. We observe that maintaining style consistency in the AgentTTS task and achieving high intelligibility in the CapTTS-SE task remain particularly challenging. In particular, models achieve good WER results but perform poorly on IMOS in the CapTTS-SE task, indicating that sound events are generated with lower quality than speech.

\section{Conclusion and Discussions}
\label{sec5}
We propose CapSpeech, a comprehensive benchmark for a series of CapTTS-related tasks. The CapSpeech dataset consists of a large collection of audio-caption pairs exhibiting diverse speaking styles. In addition, we introduce two new real-world datasets for the AgentTTS and CapTTS-SE tasks. We evaluate two strong models on the benchmark, both demonstrating high-fidelity and highly intelligible speech generation. However, we find that maintaining style consistency in the AgentTTS task and achieving high intelligibility in the CapTTS-SE task remain particularly challenging.


Some limitations of this work are language coverage and evaluation metrics.
While the design of our benchmark can be readily extended to other languages, the current datasets are limited to English.
Moreover, the style-captioned TTS tasks rely on costly and subjective human evaluations due to the absence of reliable automatic metrics. Currently, no existing understanding model can generate high-quality speech captions. However, our dataset provides a promising foundation for training such models, analogous to image-text models like CLIP \cite{DBLP:conf/icml/RadfordKHRGASAM21} and BLIP \cite{DBLP:conf/icml/0001LXH22}. In addition, the scale of CapSpeech-SEDB and CapSpeech-AgentDB is still limited, we are actively collecting more data in the future works.
 
\bibliographystyle{IEEEtran}

\bibliography{ref}

\vfill

\end{document}